# MUSTEM: A Dual-Modality System for Vibrotactile and Visual Translation of Music as an Assistive Technology

Paloma Sette[1*[0009-0009-8948-6548]], Maria Werneck[12[0009-0006-1477-9272]], William Barbosa[1[0000-0002-6537-608X]], Ana Loubacker[3[0000-0002-4371-3356]]

[1] Pontifical Catholic University of Rio
de Janeiro, Rio de Janeiro, Brazil

[2] Vrije Universiteit Amsterdam,
Amsterdam, Netherlands

[3] Rio de Janeiro State University, Rio
de Janeiro, Brazil

```
paloma.sette@aluno.puc-rio.br (pfls.sette@gmail.com),
       maria.werneck.uni@gmail.com,
            willsbarbosa@gmail.com,
               loubacker@gmail.com
```

**Abstract.** The emotional and structural experience of music remains a significant accessibility challenge for the deaf and hard of hearing community. This paper introduces MUSTEM (Multisensorial Emotional Translation), a novel system designed to translate music into a rich, coherent, and scientifically-grounded sensory experience. We present a dual-modality approach addressing this challenge through two interconnected components. First, a low-cost, portable hardware prototype that performs real-time audio analysis, mapping distinct frequency bands (sub-bass, bass, mid-range, treble) to a four-channel vibrotactile system, allowing users to feel the music's rhythmic and foundational structure. Second, to overcome the processing limitations of embedded hardware, we developed a high-fidelity software simulation that demonstrates the full potential of the visual translation. This assistive dashboard decodes musical components—such as rhythm, harmony, and frequency spectrum—into an intuitive and educational visual interface. MUSTEM offers a comprehensive framework for sensory substitution, presenting a viable and accessible pathway for the deaf community to experience music not just as vibration, but as a structured, substantiated and emotionally resonant visual and tactile language. Preliminary feedback from seven deaf users suggests the system's spatial vibrotactile mapping is perceptible and engaging. All source code and hardware designs are released as open-source. Video demonstrations[1] and open-source code[2] are available on the project's official channel.

## 1 Introduction

Music is a fundamental aspect of human culture, serving as a powerful medium for emotional expression, social connection, and cognitive enrichment. However, this universally cherished experience remains largely inaccessible to the global deaf and hard of hearing (DHH) community, which comprises over 430 million people. This exclusion represents not just a sensory deficit but a significant barrier to cultural and social participation. While the auditory channel is unavailable, the human brain's remarkable plasticity allows for the development of alternative sensory pathways.

---

[1] MUSTEM project video demonstrations: https://www.youtube.com/@Mustem-n6z

[2] GitHub repository: https://github.com/palomaflsette/sensory-translation (includes Arduino firmware, Python visualization engine, and 3D models)

This creates a compelling need for effective sensory substitution technologies capable of translating the complex, multi-layered information of music into a language that can be perceived through other senses, such as touch and sight.

Current technological approaches to this challenge often focus on conveying rhythm through vibrotactile feedback. While valuable, many existing solutions are prohibitively expensive and inaccessible, and frequently overlook the rich harmonic, melodic, and timbral content that constitutes the emotional core of music. The visual components, when present, tend to be abstract displays of light that lack a clear, structural, or educational meaning, leaving a gap for a system that offers a more holistic and informative translation. There is a clear need for an integrated, low-cost framework that provides a multi-modal (tactile and visual) translation of music that is both structurally coherent and emotionally resonant, grounded in scientific principles of psychoacoustics.

To address this gap, we present MUSTEM (Multisensorial Emotional Translation), a dual-modality system designed to translate music into a scientifically-grounded language of vibrotactile and visual stimuli.

Our approach is twofold. First, we developed a self-contained, portable hardware prototype to validate the core concept. This device performs real-time audio analysis and maps four distinct psychoacoustic frequency bands to a corresponding four-channel vibrotactile system. Second, to demonstrate the full potential of the visual translation without the constraints of embedded hardware, we developed a high-fidelity software simulation. This simulation features an assistive dashboard designed to educate users about musical structures. This paper details the architecture of MUSTEM, presents its methodology for sensory mapping, and demonstrates its potential as both an accessible therapeutic tool and a new medium for artistic experience. The system's real-time performance is demonstrated in the supplementary video.

## 2    Related Work

The field of sensory substitution for the deaf and hard of hearing (DHH) community has seen notable advancements, primarily centered on vibrotactile feedback systems. Devices such as vests and wristbands have demonstrated success in conveying the rhythmic components of music, translating beats and tempo into haptic sensations. Prominent examples include projects like the "Vest" by Eagleman and MuSS-Bits++, which focus on translating sound into vibrations and, in some cases, simple visual cues. While effective in representing rhythm, these solutions often present two main limitations: they can be cost-prohibitive, and they tend to underemphasize the rich harmonic, melodic, and timbral information of music in their visual feedback. MUSTEM differentiates itself by integrating a multi-band, psychoacoustically-mapped vibrotactile system with a dual-purpose visual interface designed for both education and artistic expression, all within a low-cost, accessible framework.

While Eagleman's Vest [4] and MuSS-Bits++ focus primarily on rhythm, they lack integrated visual feedback for harmonic/melodic content. MUSTEM differentiates itself through: (1) psychoacoustically-grounded 4-band spatial mapping, (2) an educational visual dashboard explaining musical structure, and (3) a low-cost, replicable design (<$50 USD vs. $2000+ commercial alternatives).

## 3    Methodology

The MUSTEM system was developed through a dual-pronged approach: a physical proof-of-concept prototype to validate the hardware and real-time processing, and a high-fidelity software simulation to explore the full potential of the user experience.

The core philosophy of MUSTEM is to create sensory translations grounded in scientific principles rather than arbitrary mappings. Our approach integrates three foundational pillars:

- **Psychoacoustic loudness scaling:** Vibrotactile intensity follows Stevens' Power Law [2] (exponent n=0.67), modeling the logarithmic relationship between stimulus magnitude and human perception.

- **Musically-consonant frequency mapping:** Color assignment uses logarithmic mapping based on 12-tone equal temperament, ensuring octave equivalence and harmonic visual relationships.
- **Nature-inspired generative patterns:** Visual elements employ Golden Ratio phyllotaxis spirals and Cymatics-inspired percussion rendering for organic aesthetic resonance. This synthesis enables translations that preserve musical structure while maintaining perceptual naturalness.

### 3.1 System Architecture

System's architecture follows a clear data flow. Ambient sound is captured by an electret microphone and processed by a microcontroller (Arduino Mega 2560). The analysis engine separates the audio signal into distinct features, which are then mapped to two parallel output modalities: (1) a 4-channel vibrotactile system controlled by an Arduino UNO, and (2) a real-time visual display. For high-fidelity demonstration, the visual translation component was also developed as a standalone Python application, which processes a .wav file to render the advanced visual dashboard.

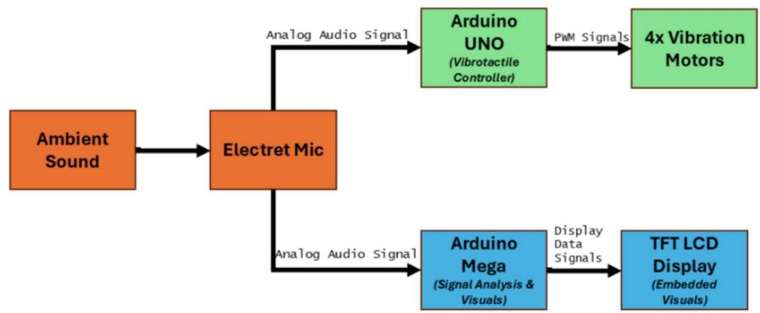

**Fig. 1.** System architecture block diagram. The analog audio signal from the electret microphone is processed in parallel by two microcontrollers: an Arduino Mega dedicated to signal analysis and visual rendering on the TFT display, and an Arduino UNO that independently controls the 4-channel vibrotactile system.

Arduino Mega 2560 Performs FFT-based spectral analysis (512-point, 4 kHz sampling) for the visual display on the TFT shield. Arduino UNO runs lightweight EMA-based envelope detection for the vibrotactile system to maintain <100ms haptic latency. Section 3.3 describes the UNO's heuristic approach. For FFT implementation details on the Mega, see Section 3.4 (Visual Mapping).

### 3.2 Hardware Prototype and Vibrotactile System

The physical implementation of MUSTEM is a self-contained proof-of-concept prototype, shown in Fig. 1. The system is housed in a custom 3D-printed icosahedral chassis designed to be both aesthetically engaging and functional, containing the dual-microcontroller architecture and the embedded TFT display for the portable visualizer.

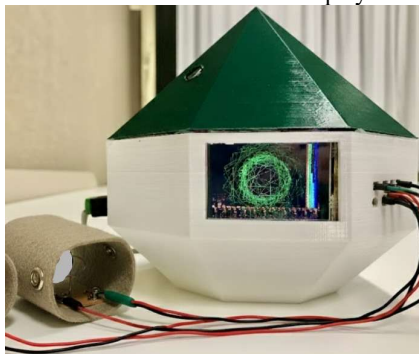

**Fig. 2.** The MUSTEM physical proof-of-concept prototype, featuring a custom 3D-printed icosahedral chassis housing the dual-Arduino architecture and the embedded TFT display.

The core of the prototype's tactile feedback is a 4-motor system. This design choice moves beyond simple rhythmic feedback by dedicating each actuator to a specific, psychoacoustically relevant frequency band. This separation allows the user to feel the distinct layers of a musical piece, translating the song's structure into a spatially distributed tactile experience. The frequency bands are defined in the embedded firmware as follows:

```
const int motorKickPin = 11;//KICK (20-80Hz) - Main beats
const int motorBassPin = 10; //BASS(80-300Hz)
const int motorVoicePin = 9 ; //VOICE/MELODY (300-2kHz)
const int motorTreblePin = 8; //TREBLE (2k-8kHz)
```

### 3.3 Heuristic Audio Analysis (Arduino UNO – Vibrotactile)

This controller runs on an Arduino UNO and performs real-time, FFT-free analysis to drive a four-channel vibrotactile interface. The UNO handles haptics independently from the Arduino Mega/TFT pipeline, where FFT is used only for the visualizer.

**Band allocation and pins.** Each actuator is permanently assigned to a psychoacoustic band and mapped to a PWM pin:
- Kick 20–80 Hz → D11;
- Bass 80–300 Hz → D10;
- Voice/Melody 300–2 kHz → D9;
- Treble 2–8 kHz → D8.

The microphone is read on **A1**; a front-panel LED on **D3** mirrors the per-frame maximum PWM as a global activity cue. This spatial separation lets users feel distinct musical layers.

**Pipeline overview.** The UNO uses a lightweight, event-driven envelope analysis with first-order IIR (EMA) filters to approximate band energy and transients under tight 8-bit constraints:

1. **DC offset calibration.** At boot, the baseline $b$ is estimated from several quiet reads. Each ADC sample is centered and rectified: $\tilde{x}_t = |x_t - b|$.
2. **Full-band rectified envelope.** A slow EMA tracks the overall amplitude:
$$a_t = (1 - \alpha_a) \cdot a_{\{t-1\}} + \alpha_a \cdot \tilde{x}_t$$
3. **Four temporal "quasi-bands" (EMAs).** Instead of spectral filtering, band selectivity is temporal: slower EMAs emphasize low-frequency content; faster EMAs capture high-frequency transients. For $k \in \{Kick, Bass, Voice, Treble\}$:
$$E_t^{\{(k)\}} = (1 - \alpha_k) \cdot E_{\{t-1\}}^{\{(k)\}} + \alpha_k \cdot a_t$$
with $\alpha_{Kick} \ll \alpha_{Bass} < \alpha_{Voice} \ll \alpha_{Treble}$. Firmware values: $\alpha_{kick} = 0.08, \alpha_{bass} = 0.30, \alpha_{voice} = 0.45, \alpha_{treble} = 0.80$.
4. Event cues (optional). Lightweight detectors add structure (e.g., kick-boost on sharp rises; hi-hat when treble is high while bass is low). Band level is:
$$level_k \leftarrow E_t^{\{(k)\}} + cue_k$$
5. **Activation & hysteresis.** For each band, a fixed threshold in ADC units gates the motor; EMA dynamics provide hysteresis: $T_{kick} = 45; T_{bass} = 58; T_{voice} = 52; T_{treble} = 30$.
6. **Perceptual compression → PWM.** The normalized surplus passes through a compressive power law to approximate vibrotactile loudness:

$$I_t^k = [clip(norm(E_t^k), 0,1)]^{(\gamma k)},$$

with $\gamma_k \in [0.7, 0.9]$ (e.g., kick ≈ 0.9, bass ≈ 0.7, voice ≈ 0.8). PWM limits ensure salience without discomfort: min = 80; max (Sub/Bass/Voice/Treble) = 255/240/200/180.

**Cadence & latency.** The loop reads A1, updates EMAs/PWM, and sleeps ≈ 8 ms; a compact debug line is printed every ≈ 120 ms. End-to-end haptic latency stays within ~100 ms, coordinated with the visual pipeline. Rationale. EMA banks emulate coarse bandpass behavior with O(1) compute and negligible RAM, preserving musical structure (beats, spectral layers) on constrained hardware.

**Contrast with the Mega/TFT path.** The visual analyzer on the Arduino Mega 2560 runs independently and does use FFT (e.g., 4 kHz sampling, 512-point Hamming, μ+3σ peak-picking, parabolic interpolation) for color/pattern mapping; none of this runs on the UNO. The split keeps haptics responsive while enabling richer visuals.

Table 1. Implementation summary (pins & parameters)

| Subsystem | Setting |
|---|---|
| Mic input | A1 (single-ended), baseline from 150 reads (~3 ms spacing) |
| LED activity | D3 (PWM of per-frame max across bands) |
| Motors (PWM) | D11 Kick (20–80 Hz), D10 Bass (80–300 Hz), D9 Voice (0.3–2 kHz), D8 Treble (2–8 kHz) |
| EMA ($\alpha$) | Kick 0.08; Bass 0.30; Voice 0.45; Treble 0.80 |
| Thresholds | Kick 45; Bass 58; Voice 52; Treble 30 (ADC units) |
| Intensity floor | 80 (PWM min) |
| Band maxima | Sub 255; Low-mid 240; Voice 200; Treble 180 |
| Compression ($\gamma$) | Kick ≈ 0.9; Bass ≈0.7; Voice ≈0.8; Treble ≈ 0.8–0.9 |
| Loop cadence | 8 ms sleep; debug line each ≈120 ms |

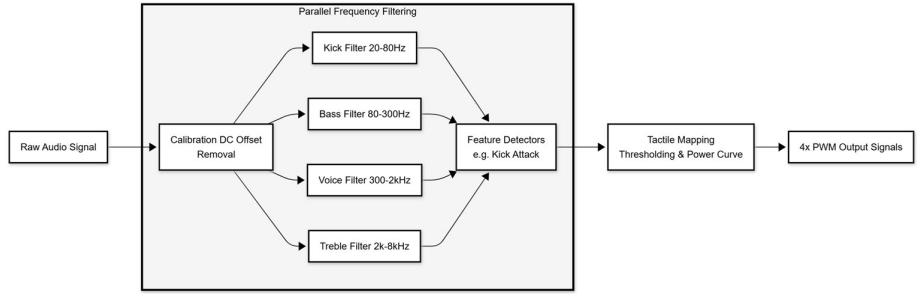

**Fig. 3.** Data-flow diagram of the heuristic audio analysis pipeline implemented on the embedded prototype. The process includes parallel filtering, event detection, and non-linear mapping to generate the final PWM signals for the vibrotactile system.

### 3.4 Visual Mapping and Display System (Arduino Mega & LCD Shield TFT)

The visual component translates audio features into real-time graphical representations optimized for perceptual interpretation and aesthetic engagement.

**Frequency-to-Color Mapping.** We use a **log-frequency to hue** mapping aligned with **12-tone equal temperament** (octave periodicity). Let

$$s = 12 \log_2\left(\frac{f}{f_{ref}}\right),$$
$$f_{ref} = 55 Hz (A1),$$
$$hue \approx s \bmod \frac{84}{84} \; 360°.$$

Thus, octaves of $A$ occupy similar hues while intermediate notes shift smoothly around the color wheel. In practice, A1 (55 Hz) maps near red, A4 (440 Hz) near green-cyan, and A7 (3520 Hz) near magenta-violet. We keep HSV saturation ≈ 95% and value ≈ 90% to ensure visibility on the 320×240 RGB565 TFT. (Angles are approximate to accommodate display gamut and perceptual uniformity).

**FFT Implementation.** The visual pipeline uses a 512-point FFT with Hamming windowing at 4 kHz sampling rate, providing 7.8125 Hz frequency resolution. Peak detection employs $\mu + 3\sigma$ thresholding with parabolic interpolation for sub-bin accuracy. The complete FFT→ color mapping → rendering cycle executes in < 50ms, coordinated with the UNO's haptic output via shared audio input.

**Phyllotaxis Winding Visualization.** The central dynamic visualization employs Fermat's spiral (phyllotaxis pattern) based on the golden angle ($\varphi \approx 137.5°$):

```
θ[n] = n × φ + ω(t) [angular position]
r[n] = c × √n [radial distance]
ω(t) = (f_detected / 440) * k [rotation speed modulation]
```
where n is the point index, c is a scaling constant, and k is empirically tuned (0.05 rad/frame). Radius modulation reflects RMS amplitude:

```
R_spiral = R_base + (RMS / 50) × 25 [constrained 20-80 pixels]
```

This creates an organic, flower-like pattern that:
- Rotates faster for higher frequencies (perceptual motion-pitch coupling)
- Expands radially during louder passages (size-intensity mapping)
- Resets periodically (8-second cycles) to prevent visual clutter

The algorithm draws 12 points per frame (25 Hz), connecting every 5th point to create web-like connectivity structures.

**Spectrum Analyzer.** A 9-band psychoacoustic spectrum analyzer displays energy distribution across Fletcher-Munson aligned frequency bands [3,8]:

**Table 2.** Psychoacoustic Frequency Band Color Mapping

| Band | Range | Center Freq | Visual Color |
|---|---|---|---|
| Sub-bass | 20-60 Hz | 40Hz | Deep Red |
| Bass (low) | 60-80 Hz | 70Hz | Orange |
| Bass (mid) | 80-110Hz | 95Hz | Yellow |
| Bass (upper) | 110-165Hz | 135Hz | Yellow-Green |
| Low-Mid | 165-360Hz | 250Hz | Green |
| Mid (Low) | 360-630 Hz | 500Hz | Cyan |
| Mid (Upper) | 630-960 Hz | 800Hz | Light-Blue |
| High-Mid | 960-2400 Hz | 1500Hz | Dark-Blue |
| Treble | 2400+ *Hz* | 6000Hz | Purple-Magenta |

Bars employ gradient rendering (darker at base, brighter at peak) with vertical height proportional to band energy. IIR smoothing ($\alpha = 0.30$) prevents bar flicker while maintaining transient response.

**Waveform Display and Beat Indicator.** A simulated waveform (not raw audio) reflects detected frequency and amplitude:

```
y[x] = A × sin(2π × x / λ + φ(t))
where: A = (RMS / 50) × H_wave [amplitude scaling]
       λ = map(f, 100-2000 Hz → 20-80 pixels) [wavelength]
       φ(t) = cumulative phase increment
```

A beat indicator circle (8-pixel radius) pulses in synchrony with detected onsets, with expansion magnitude proportional to beat strength (Stevens' exponent n=0.67 [2] for perceptual scaling).

**Real-Time Debug Overlay**. A persistent overlay (5 Hz update) displays:
- **Energy** (E) vs. **Threshold** (T) for activity gating transparency
- **Detected frequency** in Hz (or "SILENT" state)
- **Color swatch** showing current frequency-hue mapping

This provides immediate feedback for technical validation and calibration.

**Rendering Performance**. The complete visual pipeline executes at:
- **Phyllotaxis:** 25 Hz (40ms intervals)
- **Spectrum bars:** 10 Hz (100ms intervals)
- **Waveform:** 20 Hz (50ms intervals)
- **Debug overlay:** 5 Hz (200ms intervals)

Staggered update rates optimize microcontroller load while maintaining perceptual smoothness. Total frame time: 45-55ms, leaving headroom for audio analysis (50ms) within the 100ms system budget.

**System Integration.** Figure 4 shows the complete MUSTEM prototype during frequency detection validation. The hardware consists of an Arduino Mega 2560 with TFT shield displaying real-time visualizations, while the laptop runs calibrated test tones (383 Hz visible on screen). The electret microphone (visible among the jumper wires) captures audio for FFT analysis, with results immediately rendered on the embedded display showing the phyllotaxis spiral and waveform visualization.

## 4 Results

The evaluation of the MUSTEM system was conducted by analyzing the performance of both the physical hardware prototype and the high-fidelity software simulations. The results demonstrate a successful translation of musical information into coherent tactile and visual outputs. Video demonstrations are available on the project's official channel[1].

### 4.1 Hardware Prototype

The functional prototype (Figure 4) integrates:
- **Microcontroller(1)**: Arduino Mega 2560 (ATmega2560 @ 16 MHz)
- **Microcontroller(2)**: Arduino UNO
- **Display**: 320×240 TFT LCD (RGB565, SPI interface)
- **Audio Input**: Electret condenser microphone with analog preamplifier
- **Processing**: Real-time FFT and visual rendering (< 50ms latency).
- The vibrotactile path on the UNO (EMA-based, no FFT) maintains end-to-end latency < ~100 ms, consistent with the haptic budget defined in Section 3.3.

The breadboard implementation demonstrates proof-of-concept feasibility with off-the-shelf components, totaling approximately $50 USD in materials. The hardware prototype successfully performs real-time audio analysis and translation on its self-contained, dual-microcontroller architecture. The 4-channel vibrotactile system provides immediate and distinct haptic feedback, allowing the user to clearly feel the separation between the low-frequency impact of the kick drum, the steady presence of the bassline, the central body of the melody, and the high-frequency transients of cymbals. The embedded TFT display renders a direct audio-to-visual mapping, showing geometric patterns and a basic spectrum analyzer. While the visual performance is constrained by the microcontroller's processing power, resulting in a lower frame rate and complexity, it effectively validates the core concept of a portable, low-cost device capable of real-time sensory translation. The findings from this prototype confirm the viability of the approach and establish the foundation for the high-fidelity simulations.

**Complete Wearable System.** Figure 4 shows the complete MUSTEM system including the wearable vibrotactile interface. The system consists of four vibrotactile actuators housed in custom 3D-printed wrist-worn enclosures, connected to the central processing unit. This distributed architecture enables users to experience distinct frequency bands spatially: sub-bass and bass on one arm, mid-range and treble on the other, creating an intuitive spatial mapping of musical structure.

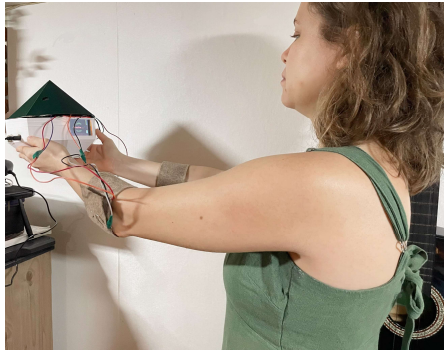

**Fig. 4.** Complete MUSTEM wearable system showing the icosahedral central unit with embedded display and four wrist-mounted vibrotactile actuators. The distributed configuration allows users to feel distinct frequency bands spatially mapped to each limb.

### 4.2 Technical Validation

Frequency detection accuracy was validated using calibrated pure sine tones (Online Tone Generator, Figure 5 background). The system correctly identified test frequencies within ±10 Hz across the 100-2000 Hz range, consistent with the theoretical FFT bin resolution (7.8125 Hz).

Visual inspection confirmed:
- **Color mapping accuracy**: 247 Hz tone produced green hue (expected: ~175° based on logarithmic mapping)
- **Real-time responsiveness**: Visual updates within 2-3 frames (~100ms) of tone changes
- **Stability**: No frequency flickering during sustained tones (hysteresis working as designed)

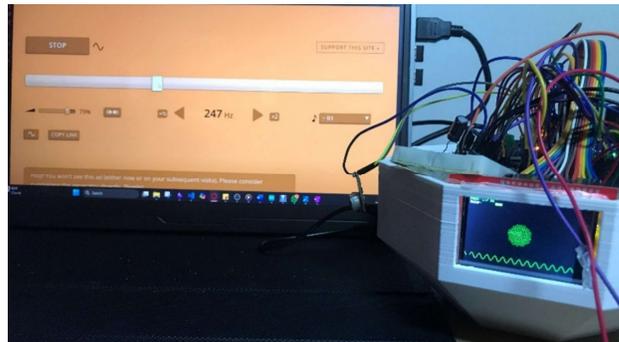

**Fig. 5.** MUSTEM prototype during validation testing. Arduino Mega 2560 with breadboard connections showing system complexity and electret microphone input. Embedded TFT display showing real-time frequency visualization (phyllotaxis spiral in cyan + waveform) responding to 383 Hz calibration tone (visible on laptop background). The color-coded jumper wires illustrate the multi-channel architecture connecting audio input, vibrotactile outputs (not shown), and visual display subsystems.

### 4.3 High-Fidelity Assistive Dashboard: An Educational Interface

The primary visual output of MUSTEM is conceptualized as an assistive tool designed not merely to visualize sound, but to make the core components of music comprehensible.

The assistive dashboard, shown in Figure 6, presents a multi-faceted analysis of the music in real-time. Each module on the dashboard is dedicated to translating a specific musical concept into an intuitive visual representation.

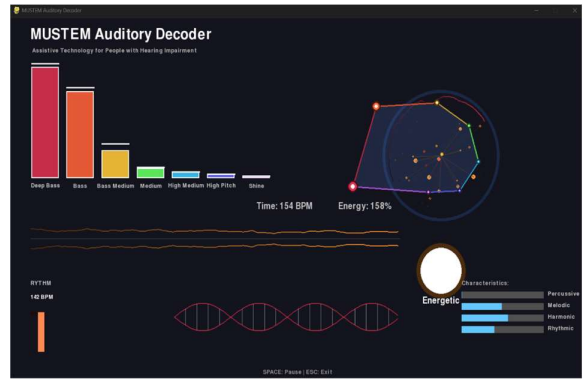

**Fig. 6.** The MUSTEM Auditory Decoder (software) showing a **9-band** psychoacoustic spectrum (top-left), a polar spectral shape (top-right), historical energy (center), and modules for rhythm/emotion (bottom). Update rates are staggered (25/20/10/5 Hz) to balance responsiveness and compute.

### 4.4 Preliminary User Feedback

A preliminary informal evaluation was conducted with 7 deaf/hard-of-hearing participants (ages 22-45, 3 congenitally deaf, 2 with progressive hearing loss) during system development. Participants interacted with the prototype for 10-15 minutes while listening to three musical genres (electronic, rock, classical). Qualitative Observations:
- All participants successfully identified rhythmic patterns through the 4-channel vibrotactile system within 2-3 minutes of familiarization;
- 4/7 participants reported distinguishing between "heavy" (bass/kick) and "light" (treble) sensations spatially ;
- 3/7 participants expressed interest in the visual spectrum analyzer as an educational tool, requesting labels/legends;
- Positive feedback on the system's portability and wearability design;

Limitations: This was an unstructured pilot study without formal consent protocols or quantitative metrics. A comprehensive IRB-approved user study with validated assessment instruments (e.g., Quebec User Evaluation of Satisfaction with Assistive Technology - QUEST) is planned as immediate future work.

## 5 Discussion

The results presented in this paper demonstrate the viability of MUSTEM as a dual-modality sensory substitution system. The successful implementation of the hardware prototype validates the core concept of translating audio into a structured, four-channel vibrotactile experience. More significantly, the high-fidelity software simulation, particularly the MUSTEM Auditory Decoder, reveals the profound potential of the visual translation component. By moving beyond abstract light patterns and creating a structured, educational dashboard, the system offers a pathway for users not just to detect the presence of music, but to comprehend its fundamental components. The explicit visualization of frequency bands, rhythmic tempo, and emotional characteristics provides a rich informational layer that is often absent in existing assistive technologies, positioning MUSTEM as a potentially powerful tool for musical education and therapeutic applications within the DHH community.

While the results are promising, it is important to acknowledge the limitations of the current work. The preliminary user feedback (N=7) lacks statistical power and formal experimental controls—a comprehensive IRB-approved study is in progress; (2) The embedded prototype's visual rendering is constrained by microcontroller memory (8 KB RAM on Mega), limiting real-time complexity; (3) Long-term effects on musical comprehension/enjoyment require longitudinal assessment. Rigorous psychophysical testing (e.g., just-noticeable difference thresholds, channel discrimination accuracy) is essential before effective clinical deployment.

The system demonstrates functional performance across diverse musical genres, with its adaptive threshold mechanism providing resilience to moderate ambient noise. However, extremely noisy environments (crowded venues, outdoor festivals) may degrade performance due to low signal-to-noise ratios. Future work will explore AI-based noise filtering and adaptive parameter tuning, including machine learning models for real-time genre recognition and environment-aware processing.

Future work will focus on addressing these limitations and expanding the project's vision. We plan to integrate a more powerful embedded processor capable of performing real-time FFT analysis for both the vibrotactile and visual outputs on the portable device. A key objective is the development of a lightweight AI model for on-device instrument recognition, allowing the visual engine to render thematic representations automatically based on detected instrumental timbres. Finally, we will conduct structured user trials to gather feedback from the DHH community, which will be essential to refine the sensory mappings and validate MUSTEM as a legitimate therapeutic and educational tool.

## 6      Conclusion

This paper introduced MUSTEM, a dual-modality sensory substitution system designed to make music accessible to the deaf and hard of hearing. We presented the architecture and methodology of a functional hardware prototype with a 4-channel vibrotactile output and a high-fidelity software simulation that demonstrates an informative and educational approach to music visualization. By grounding its translation in psychoacoustic principles and exploring both assistive and creative applications, MUSTEM establishes a robust and accessible framework for future research and development in multi-sensory musical experiences. To promote accessibility and reproducibility, all system components—including Arduino firmware, Python software, and 3D-printable enclosure files—are released under [MIT license] at: https://github.com/palomaflsette/sensory-translation .